\documentclass[traditabstract]{aa}
\usepackage{graphicx}
\usepackage{natbib}
\usepackage{txfonts}

\usepackage{booktabs}
\bibpunct{(}{)}{;}{a}{}{,}

\begin{document}

  \title{Inclusion of Horizontal Branch stars in the derivation of star formation histories of dwarf galaxies: the Carina dSph}
  
  \author{Alessandro Savino\inst{1,2}, Maurizio Salaris\inst{1} \and Eline Tolstoy\inst{2}}

\institute{Astrophysics Research Institute, 
           Liverpool John Moores University, 
           IC2, Liverpool Science Park, 
           146 Brownlow Hill, 
           Liverpool L3 5RF, UK, \email{A.Savino@2014.ljmu.ac.uk, M.Salaris@ljmu.ac.uk} 
            \and Kapteyn Astronomical Institute, University of Groningen, Postbus 800, 9700 AV Groningen, The Netherlands
            \email{etolstoy@astro.rug.nl}  
           }

 \abstract{We present a detailed analysis of the Horizontal Branch of the Carina Dwarf Spheroidal Galaxy by means of synthetic modelling techniques, 
taking consistently into account the star formation history and metallicity evolution as determined from main sequence and red giant branch spectroscopic observations. 
We found that a range of integrated red giant branch mass loss values of 0.1-0.14~$M_{\odot}$ 
increasing with metallicity is able to reproduce the colour extension of the old Horizontal Branch. 
Nonetheless, leaving the mass loss as the only free parameter is not enough to match the detailed morphology of Carina Horizontal Branch. 
We then investigated the role played by the star formation history on the discrepancies between synthetic and observed Horizontal Branches. 
We derived a ``toy'' bursty star formation history that reproduces well observed horizontal branch star counts, and also   
matches qualitatively the red giant and the turn off regions. 
This bursty star formation history is made of a subset of age and [M/H] components of the star formation history based on turn off and red giants only, and 
entails four separate bursts of star formation of different strenghts,  
centred at 2, 5, 8.6 and 11.5 Gyr, respectively, with mean [M/H] decreasing from $\sim -$1.7 
to $\sim -$2.2 when the age of the burst increases, and a Gaussian spread of $\sigma$ 0.1~dex around these mean values. 
The comparison between the metallicity distribution function of our bursty star formation history  
and the one measured from the infrared CaT feature using a CaT-[Fe/H] calibration shows a qualitative agreement, 
once taken into account the range of [Ca/Fe] abundances measured in a sample of Carina stars,   
that causes a bias of the derived [Fe/H] distribution toward too low values.
In conclusion, we have shown how the information contained within the horizontal branch of Carina (and dwarf galaxies in general) can  
be extracted and interpreted to refine the star formation history derived from red giants and turn off stars only. 
}
\authorrunning{A. Savino et al.}
\titlerunning{Horizontal Branch stars and star formation history of Carina}
  \maketitle


\section{Introduction}

Dwarf galaxies (DGs) play a major role in modern astrophysics as they are believed to be the building blocks of the
process of galaxy formation. Therefore DGs constitute a sort of fossil record of the formation epoch of the cosmic structures,
and the determination of their star formation history (SFH) is crucial to understand the mechanisms of galaxy formation and early
evolution. 

Detailed SFHs of DGs can only be determined in the Local Group where they can be resolved into individual stars
down to the oldest main sequence (MS). These
determinations are usually based on the theoretical interpretation (via stellar evolution models and isochrones) of 
observed colour magnitude diagrams (CMDs) and, when available, spectroscopic heavy-element abundances, typically of
the red giant branch (RGB) populations.
The primary age indicators for these populations are found in the main sequence turn-off (TO) region of the CMD,
which is located at faint magnitudes for the oldest populations, and is very sensitive to photometric errors. 

The horizontal branch (HB) is routinely neglected in SFH determinations of DGs, in spite
of its brightness compared to the old TO, and the extreme sensitivity --in
terms of colour and brightness distribution-- to the mass and metallicity distribution of the parent stars.
The reason is that the interpretation of
the HB morphology in potentially simpler populations like Galactic globular clusters is problematic. 
Numerous studies of Galactic globular clusters (GCs) have shown that age and metallicity
alone cannot account for the mean colour and extension of the HB \citep[e.g.,][]{Catelan09,Dotter10,Gratton10}.
A major difficulty is that stars arriving on the Zero Age HB (ZAHB) have lost mass during the previous RGB phase,
and to date it is still impossible to predict from
first principles the amount of mass lost by RGB stars.
This issue is further complicated by the currently well established presence in individual GCs
of multiple populations of stars with enhanced helium abundances at fixed metallicity  
\citep[e.g.][and references therein]{Gratton12}, which affect both the colour and magnitude of the HB.

Despite these complications, and the presence of He-enhanced populations at fixed metallicity may well be
just a feature of GCs, HB stars contain a wealth of information to constrain the star formation rate   
and metallicity evolution of DGs at earliest times. 
In \citet{Salaris13} we presented the first detailed simulation of the HB 
of the Sculptor Dwarf Spheroidal (dSph) galaxy by means of synthetic modelling techniques, taking into account the 
SFH and metallicity evolution determined from the MS and RGB spectroscopic observations. We found
that the number count distribution along the observed HB could be reproduced with a simple
mass loss law \citep[that agrees very closely with the determinations by][for a sample of GCs]{origlia14},
and that there is no excess of bright stars that require He-enhanced populations.

The purpose of the present work is to investigate, by means of the same synthetic HB modelling as for Sculptor,
the HB populations of another DG belonging to the Local Group, the Carina dSph galaxy.
This galaxy has a SFH very different from that of Sculptor, showing multiple star formation episodes 
separated in time and in the amount of stellar mass involved. Numerous studies have
probed a broad range of properties, through deep photometric investigations \citep[e.g.][]{Smecker96, Monelli03, Bono10};
spectroscopic analysis, both at medium \citep[e.g.][]{Smecker99, Koch06, Helmi06} and high \citep[e.g.][]{Shetrone03,
  Koch08,Fabrizio12, Lemasle12, Venn12} resolution; variable star characterisation \citep[e.g.][]{Saha86, Mateo98, Dallora03, Coppola13}
and SFH analysis \citep[e.g.][]{Pasetto11, Small13, deBoer14}.
These detailed works provide us with the information needed to make a meaningful comparison between 
synthetic and observed HB populations. 
In particular, our analysis will provide strong additional constraints on the controversial issue of the galaxy metallicity distribution function (MDF) as determined spectroscopically
and photometrically from RGB stars \citep[see, e.g.,][for different conclusions about
the consistency between spectroscopic and photometric MDFs]{Bono10, vadb15}.
  
This work is structured as follows: we present the data set used for our investigations in \S2; compare the synthetic and
observed HBs in \S3; discuss the impact of our results on the SFH in \S4 while a summary is presented in \S5.
 
\section{Data}
\begin{figure}
\centering
\includegraphics[scale=0.4500]{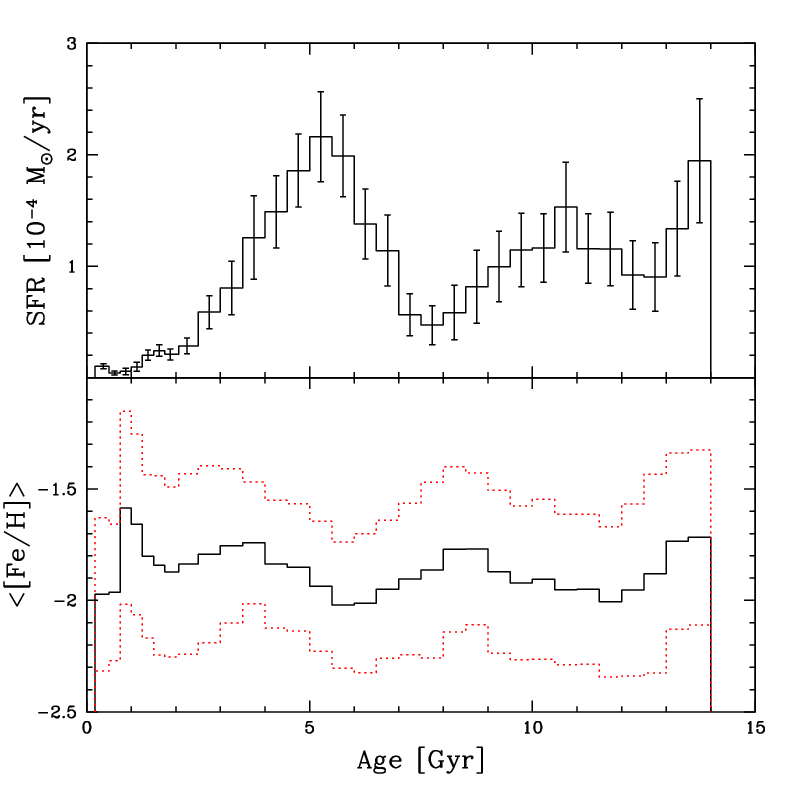}
\caption{ Carina dSph. \textit{ Upper panel:} Star Formation Rate as a function of time from the adopted SFH. \textit{Lower panel:} Weighted mean metallicity as a function of time. The red dotted histograms show the $1\sigma$ dispersion.}
\label{SFH}
\end{figure}

\begin{figure}
\centering
\includegraphics[scale=.4500]{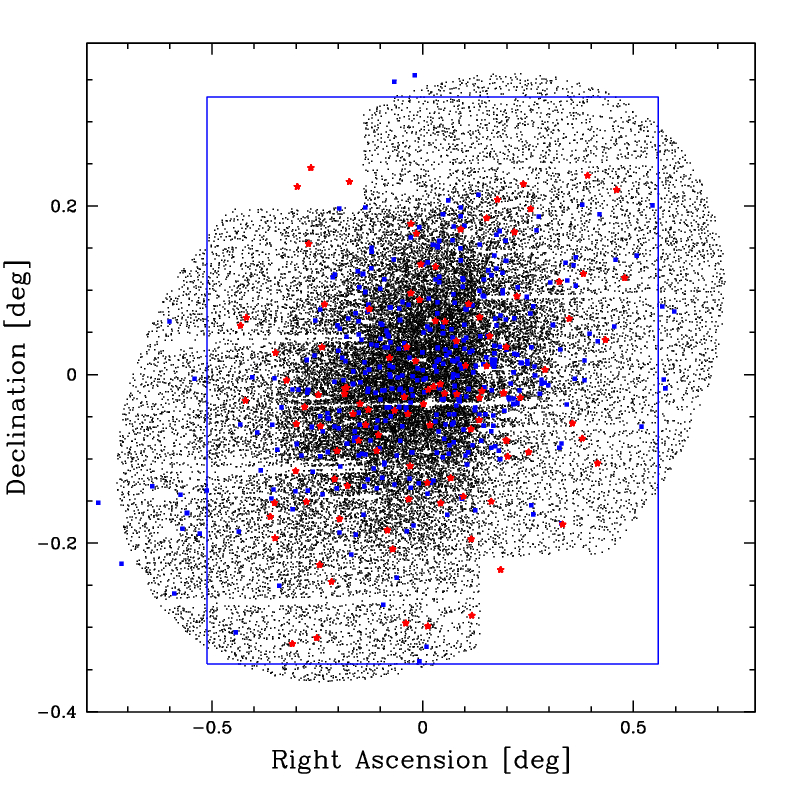}
\caption{Carina dSph. Comparison between the different fields of our data-set. The black dots represent stars of \cite{deBoer14} photometric catalogue inside one tidal radius. The blue rectangle is the field of the \cite{Bono10} catalogue. The red stars mark the position of RR Lyrae stars while the blue squares are the stars with spectroscopic measurements \citep{Helmi06}. 
The zero point of the horizontal and vertical axis is set to RA = $6h\,41m\,36.6s$ and Dec =$ -50^{\circ}\,57'\,58''$ (J2000).}
\label{Field}
\end{figure}

For the computation of our synthetic HB we employed, as reference SFH, that determined by \cite{deBoer14} (hereafter dB14). Among the several SFHs in literature, we chose the dB14 solution because it is the only one that combines the photometric modelling of the CMD with spectroscopic information about the metallicity distribution. Due to the very high sensitivity of the HB morphology to metal content, this approach is preferable.

 For the sake of homogeneity we adopted the dB14 solution obtained with the BaSTI evolutionary tracks \citep{Basti}, which are the same ones used in our synthetic HB 
calculations. Carina photometry has been divided by dB14 into three concentrical annuli inside the tidal radius of the galaxy, plus a fourth field external to the tidal radius. For each field, an independent SFH has been computed. The solution gives the star formation rate in a grid of age and [Fe/H] bins, 
with also an estimate of $[\alpha$/Fe] in each bin. The SFH used in our synthetic HB calculations refers to the sum of the annuli inside the  tidal radius.
Our Synthetic HB simulation with this SFH will be denoted as \textsl{reference} simulation.

Figure~\ref{SFH} shows star formation rate and mean stellar [Fe/H] as a function of time. There are at least two major epochs of star formation at old and intermediate ages, ranging between 9-14 and 3-7 Gyr respectively, while the mean metallicity remains in the relatively  narrow interval $-2<[Fe/H]<-1.6$.

We compare the synthetic CMDs with the photometric data from \cite{Bono10} (hereafter B10), which include 4152 CCD images acquired between December 1992 and January 2005. Although the use of dB14 photometry would have granted a perfect match between the stellar population sampled and the SFH, we chose to employ a different data-set because of the several advantages it involves.

 First, the smaller photometric errors in the B10 CMDs, of the order of $\sim 0.004$ mag at the HB level, allow for a more robust comparison between observed and synthetic CMDs. 
In addition, the same field has been the subject of a deep variable star search (Coppola et al. submitted). Furthermore, as the B10 photometric catalogue merges observations taken at different epochs, 
different random phases in different frames, for a given variable star, tend to be averaged. The colour and magnitude of possible undetected RR Lyrae in B10 will therefore 
be closer to the intrinsic mean values than it is for single epoch data-set.
 
 \begin{figure}
\centering
\includegraphics[scale=.4500]{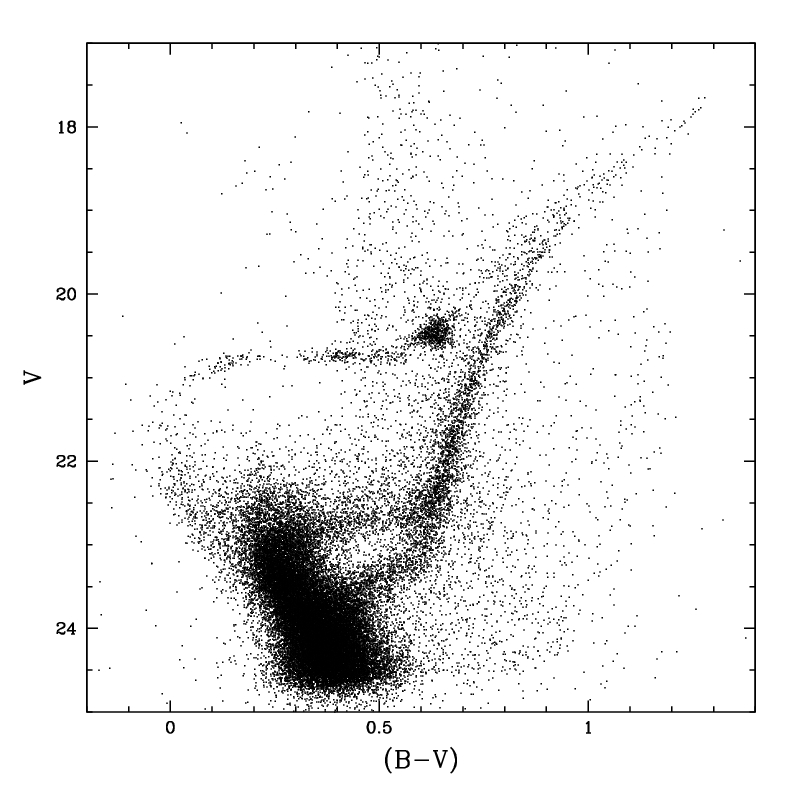}
\caption{CMD of the Carina dsph from the B10 photometric set}
\label{Phot}
\end{figure}
 
 Finally, the photometric data-set from dB14 retains the contamination by foreground stars of the Milky Way, as it is statistically taken into account at later stages during the SFH determination. Since Carina is a fairly diffuse stellar system at relatively low Galactic latititude, the number of foreground stars is large and, because the HB is much less populated compared to other evolutionary stages, the contamination can be a serious problem. The B10 catalogue, instead, has been carefully cleaned of Milky Way stars and unresolved galaxies.

Fig.~\ref{Field} shows the comparison between the dB14 field inside the tidal radius and the B10 field. 

In theory, a spatial gradient in the stellar  population properties may affect our comparison, due to the different areas sampled. Nonetheless, as the difference in the spatial coverage is a minor fraction of the total sampled area, and since the star counts are dominated by the central region of the galaxy, we expect this effect to be quite small.
It should be also noted that the results presented in this paper still hold when the synthetic CMD is compared with the dB14 photometric catalogue.

For the variable star modelling, we used the catalogue from Coppola et al. (2015, submitted), which contains information about spatial position, intrinsic mean colour and magnitude, as well as pulsation properties of RR Lyrae stars. Using the stars' coordinates, we removed variables caught at random phase in our photometry, and replaced them with their intrinsic position on the CMD. We used the period distribution to add additional observational constraints to our HB simulations.

A glance at the Carina $(V, B-V)$ CMD (Fig.~\ref{Phot}) reveals the complexity of its stellar population. Two distinct MS Turn-Offs can be clearly seen, with distinct SGBs merging in a very narrow RGB. The accepted scenario envisages Carina as a system that has undergone two or more major events of star formation, separated by a period of several Gyr \citep{Smecker96, Monelli03}. In addition, a very young population is probably present, as suggested by the presence of a blue plume above the MS and by the detection of several Anomalous Cepheids \citep{Monelli03}. The narrow RGB, coupled with the observed broad metallicity distribution \citep{Koch06, Helmi06}, suggests a conspiracy among age, metal content and alpha element abundances that leads to all stars having a very similar colour on the RGB.

The burstiness of Carina SFH can be seen in the helium burning loci as well, with an old extended Horizontal Branch that is clearly detached from a younger, more populated Red Clump 
 (RC). The discretness of the Carina stellar populations is very helpful since it allows us to simplify our analysis and model each component separately.

\section{Synthetic Horizontal Branch modelling}

We computed synthetic HB models with the code developed and fully described by \cite{Salaris13} together with the BaSTI library of scaled solar evolutionary tracks \citep{Basti}. 
The use of scaled solar models with the same total metallicity [M/H] 
of Carina stars is justified, since $\alpha$-enhanced evolutionary tracks closely mimic scaled solar ones with the same total metallicity Z, in the 
metallicity regime of this galaxy \citep[see, i.e.,][]{Salaris93}.

 We adapted the code, to allow for the large mass range on the Carina HB reaching $\sim\!1.4 \,M_\odot$. Higher masses correspond to a population younger than $1\,Gyr$ (at Carina's typical metallicity) and  can be ignored in our analysis, as their magnitudes in the helium burning phase are considerably brighter than the RC and are anyway a small contribution to the SFH of Carina.

Briefly, for each each bin of the input SFH, a number of synthetic stars is generated spanning the whole range of age and metallicity inside the bin 
(employing a uniform probability within both age and metallicity bins), 
and with a value of [$\alpha$/Fe] given for that bin. For each star, the 
corresponding evolutionary track is computed interpolating in mass and metallicity the tracks from the BaSTI grid. If the age of a star is greater than the age at the RGB tip, a 
specific amount of mass is removed and the position on the corresponding HB track is determined.

Once the position of a star in the CMD is determined, its colour and magnitude are perturbed with a magnitude dependent Gaussian photometric error, 
as provided with B10 photometry. 
We then applied to the resultant synthetic CMD a reddening of $E(B-V)=0.06$ \citep{Schlegel98} and a distance modulus of $(m-M)_0=20.11$, as used in dB14.
 Unless specified differently, we populated our synthetic CMDs with a considerably higher number of stars with respect to the observed CMD of Carina in order to minimize the Poisson error in the final model.

\begin{figure}
\centering
\includegraphics[scale=.4500]{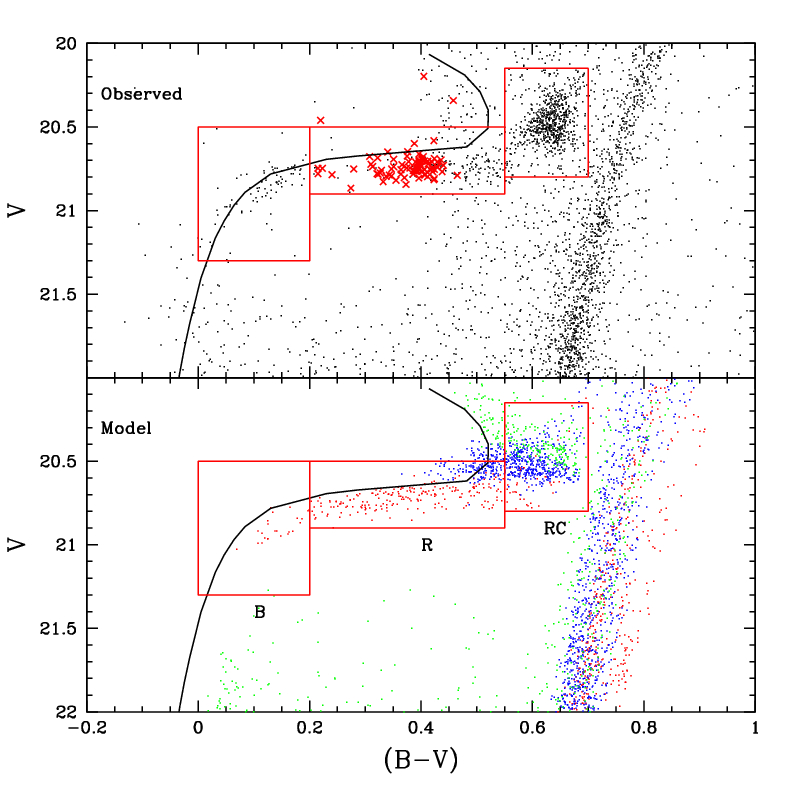}
\caption{\textit{Upper panel:} Observed CMD of the HB region of the Carina dSph from B10. Red crosses mark the position of RR Lyrae stars. The red boxes are those chosen  for detailed comparison. The solid black line represent the ZAHB, extending up to a mass of $1.45 M_\odot$, for a metallicity of Z=0.0001. 
\textit{Lower panel:} The synthetic HB for Carina. The red, blue and green dots mark stars with ages $t>7 Gyr$, $4 Gyr<t<7 Gyr$ and $t<4 Gyr$ respectively.
}
\label{bcloud}
\end{figure}

\begin{table}
\caption{The integrated RGB mass loss ($\Delta M^{RGB}$) prescription as originally determined for the Sculptor dSph 
and used for the older Carina population, together with the one used for the intermediate age population in \S~\ref{SFHcap}.}
\begin{tabular} {lll}
\toprule
 Metallicity range&$\Delta M^{RGB}_{Sculptor}$& $\Delta M^{RGB}_{\,t<7Gyr}$ \\
 &$M_\odot$&$M_\odot$\\
 \midrule
$[M/H]<-1.8$ & 0.095 & 0.048\\
$-1.8<[M/H]<-1.3$ & 0.14 & 0.07 \\
$-1.3<[M/H]$& 0.16 & 0.08\\
\bottomrule
\end{tabular}
\label{tab:ML}
\end{table}
The comparison of the model and the observed HB is a two-step process. First, an initial analysis is made by eye to see whether the colour extension and the main features of the HB are recovered. Then, after rescaling the total number of HB stars in the synthetic sample to the observed counterpart, 
we compare star counts and the mean colour and magnitude inside three boxes that encompass the RC (hereafter box RC), 
the red HB plus the RR Lyrae Instability Strip (IS -- hereafter box R) and the blue HB respectively (hereafter box B --see Fig.~\ref{bcloud}). 
If the star counts are reproduced within one $\sigma$ Poisson uncertainty and the 
difference in the mean photometric properties are within $\pm0.01$~mag, we consider the synthetic model to be a good match.

We stress that we take into account only the uncertanties arising from the Poisson distribution of the star counts. Another source of uncertainties is represented by the errorbar associated with the star formation rate in every bin of our reference SFH. Unfortunately, including this uncertainty in our analysis is problematic, as the errors are not independent from each other. Indeed, the conservation of the total number of star and of the density distribution across the CMD required by the SFH fitting procedure introduces a correlation among the individuals errorbars, so a realistic modelling of the resulting uncertainty on the model HB star counts is unfeasible without the covariance matrix of the SFH solution that is not provided by dB14. 
Undoubtedly, the issue of a proper inclusion of the SFH errors deserves to be addressed in future works.

Clearly, theoretical HB models have also intrinsic uncertainties, that affect the predicted HB star counts as a function of colour and magnitude.
The main source of uncertainty is related to the treatment of the He-core convective mixing \citep[see, e.g.,][and references therein]{cs, Gabriel14, Spruit15} that affect 
evolutionary timescales, luminosities and morphology of the HB tracks in the CMD. The treatment of core mixing in the adopted BaSTI models includes semiconvection, as 
described in \citet{Basti}, and allows to reproduce the so-called $R_2$-parameter, defined as the number ratio of asymptotic giant branch to HB stars, 
measured in a sample of galactic Globular clusters, that is very sensitive to the treatment of core mixing during the HB phase \citep[see][for a thorough discussion]{Cassisi03}.

We also computed the pulsation period of synthetic stars inside the IS and compared them with the observed period distribution. We employed both the IS boundaries and the pulsational equation from \cite{DiCriscienzo04}, assuming a mixing length $ml=1.5\,H_p$.  This choice reasonably matches the IS boundaries as inferred from the RR Lyrae colour distribution. The First Overtone periods ($P_{FO}$) are related to the Fundamental ($P_F$) by the relation $\log{P_F}=\log{P_{FO}}+0.13$ \citep{DiCriscienzo04}.

 As regards the stars in the so called OR zone, in which a RR Lyrae can pulse F, FO or double mode, we treated them as F pulsators. This is, in principle, a rough approximation but, given the small fraction of FO pulsators in the RR Lyrae sample ($\sim14\%$), we expect the number of FO pulsators in the OR zone to be very small.

\begin{figure}
\centering
\includegraphics[scale=.4500]{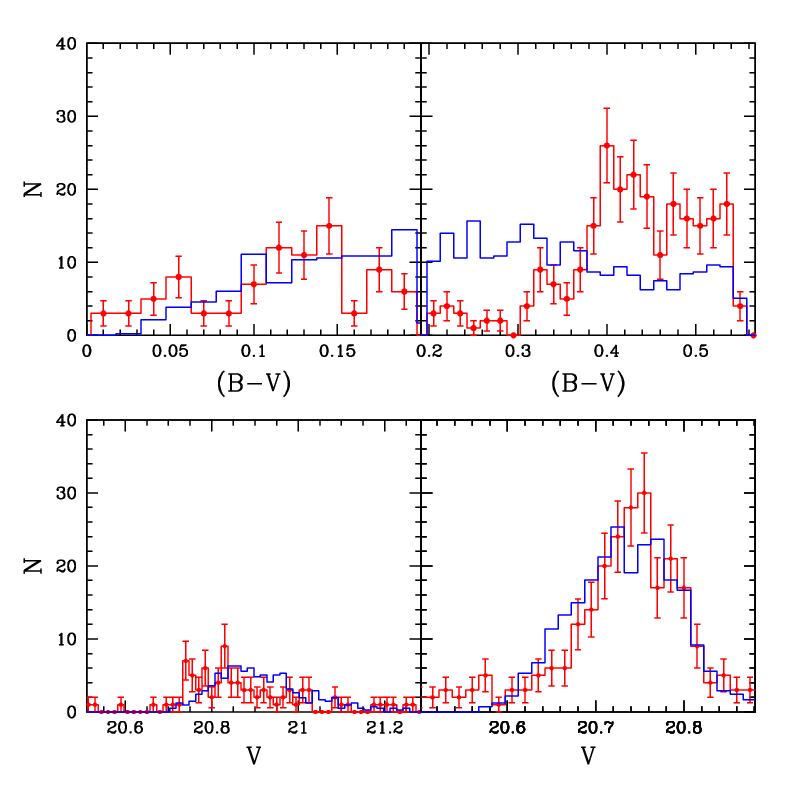}
\caption{\textit{Upper panel}: Observed (red) and synthetic (blue) star counts as a function of $(B-V)$ colour in boxes B and R (left and right respectively). 
The bin size is 0.015 mag. Poisson errors on the observed star counts are also displayed. \textit{Lower panel}: As the upper panel, but for the V magnitude.
}
\label{histold}
\end{figure}

\subsection{Results with the reference simulation}

As a starting point, we employed the mass loss law as determined by \cite{Salaris13} for the Sculptor dSph, that gives an increasing value for the integrated RGB 
mass loss with increasing metallicity, as described in Table~\ref{tab:ML}. Figure~\ref{bcloud} shows the HB of Carina as well as the synthetic model computed from the SFH. Different colours mark different age ranges. For the sake of comparison, we show a synthetic HB with approximately the same number of stars as observed in Carina's HB.

  \begin{figure}
\centering
\includegraphics[scale=.4500]{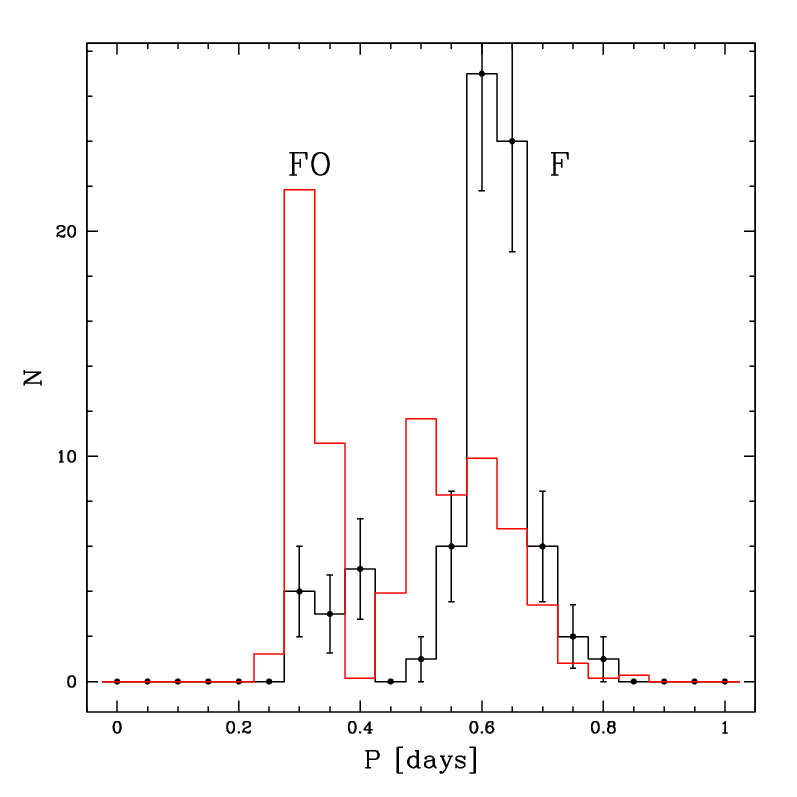}
\caption{Observed (black) and synthetic (red) period distribution for RR Lyrae stars. The bin size is 0.05 day. Poisson errors on the observed counts are also displayed.
}
\label{rrold}
\end{figure}

We can notice two major differences between synthetic and observed HB.
Firstly, what is manifestly poorly reproduced is the morphology of the RC, which is more disperse than the observed one, and with a ``blue cloud'' of stars that blend with the red end of the old HB.
   Some problems are not completely unexpected since Sculptor, for which the mass loss was determined to be dependent only on the metallicity, hosts only an old stellar population. Given the mass range of the stars on the RC, their lifetime on the RGB is considerably shorter with respect to the old population. We can, therefore, naively expect the total mass loss to be different, plausibly lower for this population. We, thus, tried to modify the mass loss for the younger population, to reproduce the observations. However, in this way, the problem could only be mitigated but not totally solved.
  
   This can be explained as follows: the RC ``blue cloud'' above the HB is primarily composed of stars with $Z\leq0.0002$. The black solid line in Fig.~\ref{bcloud} shows the Zero Age Horizontal Branch (ZAHB), extending to a mass of $1.45\,M_{\odot}$, for $Z=0.0001$, which is typical of the bulk of the metal poor population in Carina. The position on the ZAHB becomes redder as the mass increases until, over a certain mass threshold, the ZAHB turns toward bluer colours, without merging with the RC. Because of this, stars of that metallicity will always be bluer than the observed RC colour, regardless of the mass loss.
   
Given that the synthetic HB is uniquely determined by the mass loss law and the input SFH, this problem is an indication that the properties of Carina stellar population recovered by the dB14 SFH model do not match exactly the true SFH.

Considering now the HB within boxes B and R (the old HB, that comprises synthetic stars with age $t\geq7Gyr$, displayed as red dots in Fig.~\ref{bcloud}), we can see that the total 
colour extension is nicely reproduced. This implies that, given the input SFH, the true mass loss for the old population cannot be drastically different from that inferred for the Sculptor dSph.
Figure~\ref{histold} compares the colour and magnitude distributions of observed and synthetic HBs. As previously stated, the Sculptor-like mass loss law well reproduces the colour extension of the HB. Furthermore, the relative star counts inside the two boxes are reproduced within one sigma. 

   \begin{figure}
\centering
\includegraphics[scale=.4500]{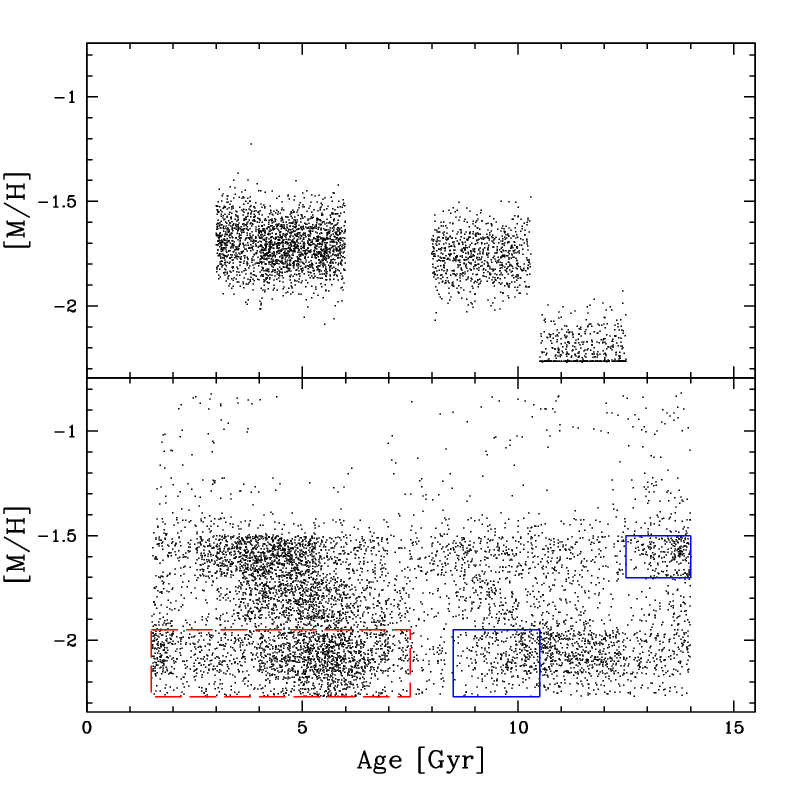}
\caption{Distribution of our HB synthetic stars in the age-metallicity plane. \textit{Upper panel:} Synthetic generated employng our toy model SFH. \textit{Lower panel:} Synthetic generated employing the dB14 SFH.
}
\label{sfhcomp}
\end{figure}

  As for the mean colour and magnitude, the fit is not perfect. In box B, the computed mean colour differs by about 0.01 mag from the observed one, while the difference in magnitudes is of the order of 0.03 mag. It should be said that, in this region of the CMD star counts are low and a few stars can significantly alter the mean colour and magnitude. 
The synthetic distributions look very consistent within the observed error bars, suggesting that the difference in the means might be due to stochasticity.
  
  In box R the V magnitude mean value and the overall distribution are remarkably well reproduced. The difference between observed and synthetic mean V is of the order of the photometric uncertainty. 
What are strikingly different are the star counts as a function of colour. The mean $(B-V)$ colour of our synthetic stars differs by more than 0.05 mag from the observed one. This can be explained by looking at the colour distribution of the two populations. The observed star counts drop for colours bluer than 0.4, and tend to increase again toward the blue HB. This ``gap'' around $(B-V)=0.3$ is missing in the synthetic HB which is uniformly populated.

We highlight here, that the procedure followed by B10 to select Carina member stars is less efficient for intermediate colour objects ($0.4<(B-V)<0.6$) and a residual amount of field stars is left in the CMD, as it can be seen in Fig.~\ref{Phot}. It is interesting to note that this contamination affects the star counts in box R and may be partly responsible for some of the observed stars at the red side of the gap, with $(B-V)>0.4$. A solid way to deal with this effect is hard to find, but we estimated that effect on the colour distribution to be of the order of 15-20\% of the typical star count. We conclude that, despite the contamination, the observed ``gap'' feature is true.
  
    \begin{figure}
\centering
\includegraphics[scale=.4500]{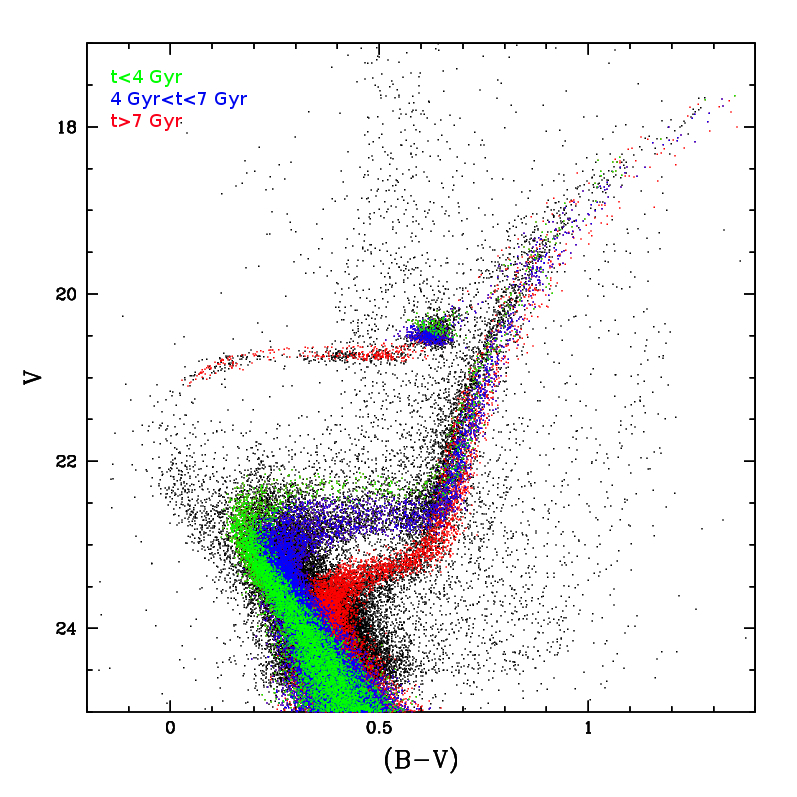}
\caption{Observed CMD of Carina (black dots) superimposed to the synthetic CMD computed from our toy SFH. Colour coding is the same as Fig.~\ref{bcloud}.
}
\label{cmdtoy}
\end{figure}

\begin{table}
\caption{Age, metallicity and normalized Star Formation Rate of the toy SFH (see text for details).}
\begin{tabular} {lllll}
\toprule
 &$t_{min}$ & $t_{max}$ & $<[M/H]>$ & SFR \\
 &Gyr&Gyr&&\\
 \midrule
Burst 1 & 3.0 & 4.0 & $-$1.69 & 0.12\\
Burst 2 & 4.0 & 6.0  & $-$1.71 & 0.39\\
Burst 3 & 8.0 & 10.3  & $-$1.77 & 0.29\\
Burst 4 & 10.5 & 12.5 & $-$2.23 & 0.2\\
\bottomrule
\end{tabular}
\label{tab:SFH}
\end{table}
  
The discrepancy in the distributions between model and data in the red box, which also includes the IS, can also be seen in Fig.~\ref{rrold}, which shows the period distributions for the RR Lyrae pulsators. The observed RR Lyrae population has a period distribution strongly peaked around $P\sim0.6\,d$ and is mainly composed of F pulsators, whereas the synthetic population has a much broader F period distribution and tends to strongly overestimate the FO population.

This can be understood by looking at the relation employed for computing the periods. The driver of the pulsational period is the effective temperature.  The periods and the colour distribution are thus closely related. Both the lack of FO pulsators and of F short periods is explained by the observed ``gap'' in the stellar distribution at the level of the IS. Our synthetic, more homogeneously populated, HB naturally covers a broader range of periods and predicts too many stars in the FO zone of the IS.

To overcome this problem we tried to change the mass loss law. An analysis of age and metallicity distributions along the HB revealed that our synthetic IS is populated by a broad range of ages and metal contents. Consequently, changing the amount of mass loss for a single value or a small range of metallicities (or ages) cannot create such a well-defined gap. On the other hand, changing the whole mass loss law will considerably affect the colour extension of the blue HB, which is extremely sensitive to small mass changes. The conclusion is that the observed gap in the HB cannot be reproduced by varying the mass loss law, except by adopting a very fine-tuned ad hoc dependence on both age and metallicity, which we see as physically hard to justify.

We have therefore investigated the SFH as a possible cause of the discrepancies encountered, to assess the additional constraints that can be made on the Carina dSph SFH using the HB modelling.

\section{Implications for the Star Formation History} \label{SFHcap}

  \begin{figure}
\centering
\includegraphics[scale=.4500]{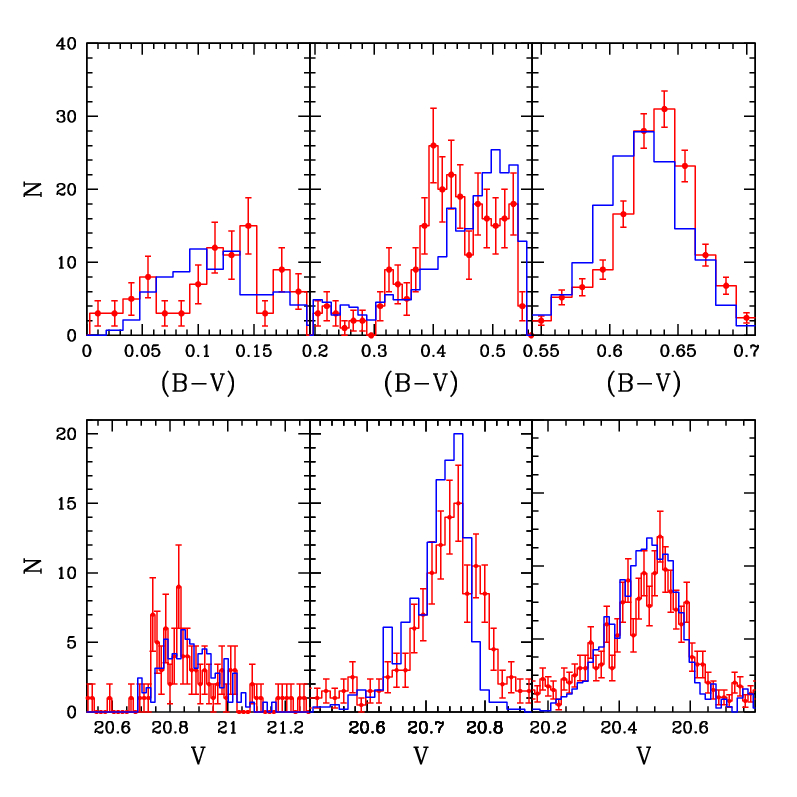}
\caption{Same plot as Fig.~\ref{histold} except for our toy SFH. The rightmost panel represents the RC box. The scale on the ordinate axis is set to 2, 2 and 10 counts per tickmark for the three upper panels (from left to right) and 1, 2 and 5 counts per tickmark for the lower panels. 
}
\label{histtoy}
\end{figure}

A complete characterization of the SFH based on the HB is not feasible, due to the uncertainty on the RGB mass loss. Nonetheless, as seen in the previous section, there are issues in our modelling, i.e. the presence in our synthetic CMD of the RC blue cloud and the missing HB gap around $(B-V)\sim0.3$, that cannot be overcome by simply changing the mass loss. 
The only other option to solve this problem should therefore be connected to the SFH solution we are employing.

We next investigated which component of the SFH could be responsible for these problems. The lower panel of Fig.~\ref{sfhcomp} shows the distribution of our synthetic HB stars in the age-metallicity plane. The region of our synthetic HB wich correspond to the observed gap  has been found to be mainly populated by two components which are enclosed within the two blue solid boxes: a group of metal poor stars with ages between 9 and 10 Gyr and a group of very old, more metal rich stars, with $[M/H]\sim-1.6$. In particular, we note that the presence of a population of old metal rich stars, is also principally responsible for the broad synthetic RGB observed in Fig.~\ref{bcloud}, as these stars populate its reddest part.

On the other hand, the RC blue cloud feature is caused, as noted in the previous section, by stars belonging to the intermediate population and with $[M/H]\lesssim -1.9$ (red dashed box in the figure).

 \begin{figure}
\centering
\includegraphics[scale=.4500]{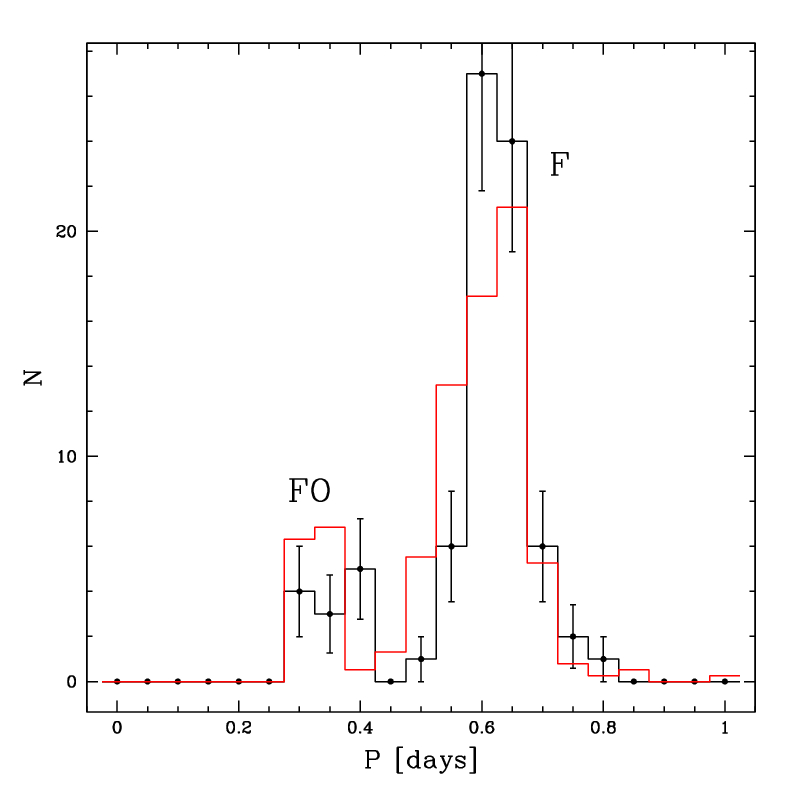}
\caption{Same as Fig.~\ref{rrold} but for our toy SFH.
}
\label{rrtoy}
\end{figure}

This finding highlights how the HB morphology can point out ``forbidden'' ranges of age and metal content that would result in unobserved features. It also suggests that the input SFH has too wide a distribution in age and metallicity, whereas Carina's true SFH is more confined to specific regions of the age-metallicity plane. 
To verify this hypothesis we built a ``toy'' bursty SFH. We tried to reproduce a synthetic HB as similar as possible to the observed one, trying at the same time to reproduce qualitatively the RGB and the TO region. The best fit model is composed of four separate bursts whose properties  
are summarized in Table~\ref{tab:SFH}. We employed a flat probability distribution for the ages in the intervals given in the table, and a Gaussian [M/H] distribution 
with the listed average values and $\sigma$=0.1~dex.
We adopted the Sculptor-like mass loss law for the old population. For the intermediate population, we assumed the RGB mass loss rate to be half that of the older population, to roughly take into account the shorter RGB lifetime (as an example, a $0.8 M_\odot$ red giant, typical of a 12-13 Gyr old population, takes $\sim1.4$ Gyr at $Z=0.0001$, and $\sim2.3$ Gyr at $Z=0.001$,  
to reach the RGB Tip from the TO. In contrast, the time taken by a $1 M_\odot$, typical of 5-6 Gyr old populations, is $\sim 0.7$ Gyr and $\sim 1.3$ Gyr, respectively).

 Figure~\ref{cmdtoy} shows the resulting synthetic CMD superimposed on the observed one. Our toy model roughly reproduces the morphology of the TOs and the SGBs, as well as the RGB colour and the morphology of the HB. Figure~\ref{histtoy} shows the colour and magnitude distributions inside the B, R and RC boxes in Fig.~\ref{bcloud}. The relative numbers inside each box are reproduced within one sigma, and the mean values of colour and magnitude differ by at most 0.01 mag from the observed ones. Even the detailed distribution looks consistent, although still with a few minor mismatches .

As an additional test, we looked at the RR Lyrae period distributions, as shown in Fig.~\ref{rrtoy}. The F period distribution is very peaked and looks similar to the observed one. The ratio between F and FO pulsators is qualitatively reproduced as well.

 It should be noted that this toy model is not intended to be the exact SFH solution, since we still lack strong constraints on the mass loss and we do not try to match the precise number density distribution across the whole CMD. Our purpose is to show how a more bursty SFH, in terms of age and metal content, is able to explain at the same time the morphology of the CMD and the detailed structure of the HB, and to suggest the age and duration of these bursts.

 \begin{figure}
\centering
\includegraphics[scale=.45]{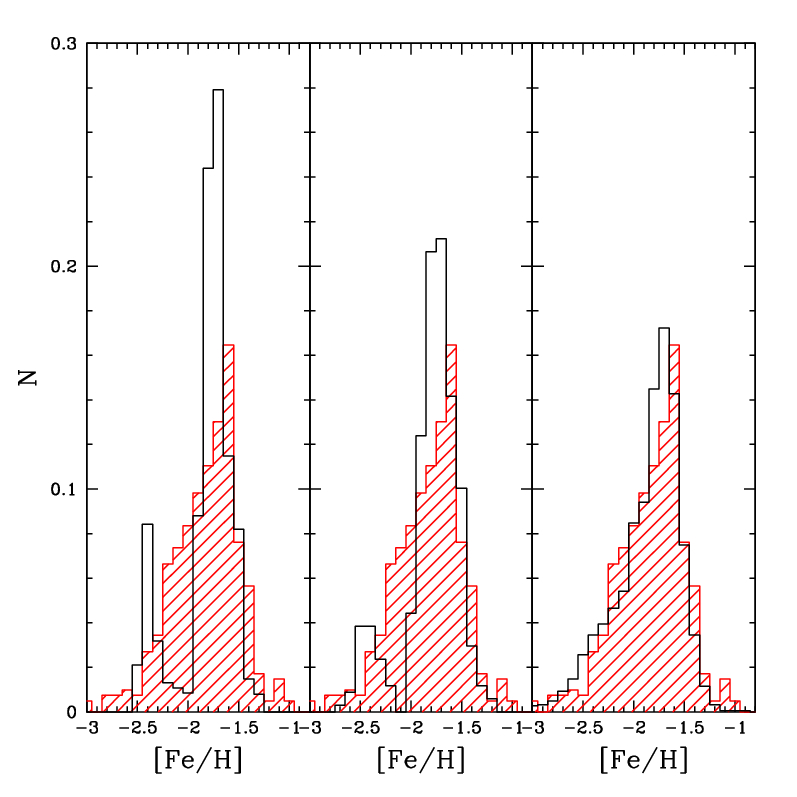}
\caption{\textit{Left panel:} Normalized MDF from Red Giants stars. The black histogram shows the MDF inferred from our toy model. 
The red shaded histogram represents the empirical distribution determined by \cite{Helmi06} and employed by dB14. \textit{Central panel:} 
as for the right panel but with the inclusion of 
intrinsic measurement errors in our MDF. \textit{Right panel:} as for the central panel but also taking into account the 
observed spread in the [Ca/Fe] values of Carina stars (see text for details).
}
\label{mdf}
\end{figure}

 The age-metallicity distribution of HB stars simulated with this model is shown in the upper panel of Fig.~\ref{sfhcomp}. The comparison with the distribution in the lower panel 
derived from the dB14 SFH allows us to identify differences and similarities between the two SFHs. The locations, in the age-metallicity plane, of all the star formation events of the toy model are roughly matched by those predicted by the dB14 solution, although the latter has a greater age spread, especially at old ages. Unsurprisingly, the major difference is the lack of metal poor stars for ages younger than 10 Gyr and of very old stars with $[M/H]\sim -1.6$.
 
 In view of this comparison, we conclude that the discrepancies between the observed HB and the synthetic helium burning population computed using the dB14 SFH are caused by two factors. First, the age resolution of the SFH naturally leads to a smooth HB, erasing any sharp substructure in the true SFH. dB14 test the age resolution of their method with a series of synthetic populations generated by a 10 Myr burst centered at different ages. Even in the best case, the recovered burst duration would have been 500 Myr, that is the width of a bin element at old ages. It can be seen from their Fig.6 that in the old age regime, the recovered SFH is a Gaussian with a dispersion of the order of 1 Gyr. This issue is due to the strong degeneracy, in the age-metallicity space, of TO and RGB, coupled with the non-negligible photometric error at that magnitude level, for a system as distant as a dwarf galaxy.
 
 As a second point, the presence of spurious components in the dB14 SFH causes a fraction of the synthetic HB population to reside in regions of the CMD which are observed to be devoid of stars. The origin of these spurious components is not clear. A possible explanation is that any difference between the spectroscopically measured MDF, which is used as a strong constraint in the SFH characterisation, and the true MDF may force the SFH determination algorithm to assign the wrong age to a fraction of stars in order to match the density distribution across the TO and the SGB regions. Regarding the Carina dSph, the problem of metallicity is particularly thorny, with photometric and spectroscopic estimates, both high and mid resolution, not always agreeing with each other \citep{Smecker96, Rizzi03, Tolstoy03, Bono10, Koch06, Lemasle12}.
 
 To investigate this point, we adopted the [$\alpha$/Fe] vs. [M/H] relation, recovered from the dB14 SFH, to compute the [Fe/H] values of our model. We computed our MDF from RGB stars, down to 3 magnitudes below the RGB tip, which is the typical selection criterion for DGs spectroscopic measurements. Inferring the metallicity distribution from the same region of the CMD is crucial, since, due to the varying evolutionary lifetime with metallicity, stars end up on the upper RGB with a different MDF with respect to the original one on the MS.
 
 The left panel of Fig.~\ref{mdf} shows our synthetic MDF compared with the measurements from \citet{Helmi06}, which have been employed in the dB14 analysis. In contrast with our sharply bimodal distribution, the measured MDF has a much broader distribution, which is unsurprising, given that the measurement errors naturally tend to smooth the underlying distribution.
 
 To check whether the two MDFs are consistent within the uncertainties, we convolved our MDF with a Gaussian error of $0.1$ dex, which is the typical uncertainty of the \citet{Helmi06} measurements. As can be seen in the middle panel of Fig.~\ref{mdf}, the bimodality of our MDF is still clearly noticeable and we conclude that the intrinsic measurements errors are not big enough to reproduce the observed MDF width which, in addition, is fairly asymmetric.
 
 It is however important to notice that this medium resolution metallicity distribution has been inferred from the CaII triplet (CaT). As \citet{Battaglia08} showed,  the equivalent width (EW) of this feature is sensitive not only to the iron abundance but also to the [Ca/H] value (as can be expected). Furthermore, the variation of other $\alpha$-element abundances, especially the [Mg/H] ratio, modifies the free electron density and, thus, the continuum opacity, indirectly affecting the measured EW. This effect has been already observed in globular clusters \citep{Mucciarelli12}. 
 
 The MDF adopted for the dB14 SFH has been evaluated using the CaT-[Fe/H] calibration by \citet{Starkenburg10}. This relation has been determined assuming a fixed metal mixture 
with [$\alpha$/Fe]=0.4 and [Ca/Fe]=0.25. As noted in the same work, at fixed iron abundance, a decrease of the [Ca/Fe] value of 0.25 dex causes a smaller EW of the CaT feature, 
mimicking a value of [Fe/H] lower by $\sim0.2$ dex
 
 Concerning Carina, several high resolution spectroscopic investigations have found a large spread in the abundance of elements like Ca and Mg, which has been explained as evidence of inhomogeneous mixing of the gas phase \citep{Shetrone03, Lemasle12, Venn12, Fabrizio15}. In particular, the bulk of the [Ca/Fe] measurements range from $\sim+0.3$ to slightly subsolar values. Using the CaT calibration without considering the variation in calcium abundance will then tend to underestimate the iron abundance. This has been already noticed by \citet{Venn12}.
 
 A precise quantitative evaluation of this uncertainty is very difficult, since it requires detailed knowledge of several quantities, like the trend of the mean [Ca/Fe] value and of its spread at varying metallicity, and the accurate EW dependence on the electron donor elements' abundance. Nevertheless, as a first approximation we took this effect into account by perturbing our MDF with an 
asymmetric uncertainty: we assumed that half of the stars had their [Fe/H] overestimated, with a dispersion of 0.1 dex, and the other half had it underestimated, with a dispersion of 0.25 dex. 
 This roughly takes into account the Carina observed distribution of [Ca/Fe]  around the fiducial [Ca/Fe]=0.25 employed in \citet{Starkenburg10} CaT-[Fe/H] calibration.
 
 The resulting distribution is shown in the right panel of Fig.~\ref{mdf}. The two distributions are in good agreement, suggesting that, at least qualitatively, an uncertainty of the order of 0.2-0.3 dex toward lower metallicities is able to produce the broad MDF we observe.
Assuming this to be the case, this issue could likely lead to spurious components in the best fit SFH, as the estimated high fraction of metal poor stars is included in the solution, with the algorithm accordingly distributing them in age to preserve the relative numbers between the two populations.

\section{Conclusions}

We have performed a detailed simulation of the HB of the Carina dSph using recent and detailed SFH estimates by dB14.  
We found that the overall colour extension of the old HB is well reproduced with integrated an RGB mass loss that ranges from 0.1 to 0.14 $M_{\odot}$ as a function of [M/H], 
in complete agreement with was has been found for the Sculptor dSph. 
This concordance hints that the mass loss 
law derived for the Sculptor dSph may reproduce the HB morphology in also other dSphs with similar metallicity.

On the other hand we found some discrepancies with the detailed colour distribution along the old HB and RC that required a modification of the input SFH by dB14.  We have then built a ``toy'' bursty SFH that reproduces well observed HB star counts in both V and $(B-V)$, and also matches qualitatively the RGB and the TO region, as well as the pulsational properties of RR Lyrae stars.

The comparison between the MDF of our bursty SFH 
and the one measured from the infrared CaT feature using the CaT-[Fe/H] calibration by \citet{Starkenburg10} shows a qualitative agreement, 
once taken into account the range of [Ca/Fe] abundances measured in a sample of Carina stars,   
that induces a bias of the derived [Fe/H] distribution toward too low values. 
Although we stress that Carina is one of the few Local Group galaxies where this issue can be so severe, extra caution should be taken when using the CaT 
to estimate [Fe/H] in system with poor or 
no constraints about the detailed metal mixture. 

In conclusion, the work we have presented shows how the information contained within the HB can, in principle, be extracted and interpreted to make predictions about the properties of the stellar population in DGs. The results on the Carina dSph illustrate that the inclusion of the detailed HB morphology in the CMD analysis is able to refine the SFH determination of resolved stellar systems. Due to the strong sensitivity of its morphology and luminosity to the age and metallicity of the parent population, the HB is a powerful benchmark to rule out ``forbidden'' ranges of age and metal content that would result in unobserved features on the CMD. This capability grants us additional resolution when pinpointing the SFH details of DGs, especially at early times.

We aim to consolidate the synthetic HB modelling technique, with the goal of making quantitative predictions on the properties of resolved stellar populations, consequently enhancing the power of CMD analysis, especially in distant systems where the old MS-TO information is hard to acces.

\begin{acknowledgements}
We thank T. de Boer, G. Bono and P. B. Stetson for providing the data used in this work and for very helpful comments and discussions.
\end{acknowledgements}

\end{document}